\documentclass[twocolumn, longauthor]{aastex631} % linenumbers

\usepackage{amsmath}
\usepackage{xspace}
\shorttitle{EELRs in post-starburst, tidal disruption event host galaxies}
\shortauthors{Wevers and French}
\begin{document}
\title{Extended emission line regions in post-starburst galaxies hosting tidal disruption events}
\correspondingauthor{Thomas Wevers}
\email{twevers@stsci.edu}
\author[0000-0002-4043-9400]{Thomas Wevers}
\affiliation{Space Telescope Science Institute, 3700 San Martin Drive, Baltimore, MD 21218, USA}
\author[0000-0002-4235-7337]{K. Decker French}
\affiliation{Department of Astronomy, University of Illinois, 1002 W. Green St., Urbana, IL 61801, USA}

\begin{abstract}
We report the discovery of an extended emission line region (EELR) in MUSE observations of Markarian 950, a nearby ($z=0.01628$) post-starburst (PSB) galaxy that hosted the tidal disruption event (TDE) iPTF--16fnl. 
The EELR requires a non-stellar ionizing continuum with a luminosity L$_{\rm ion, min} \gtrsim 10^{43}$ erg s$^{-1}$, inconsistent with the current weak state (L$_{\rm IR,AGN} < 2.5 \times 10^{42}$ erg s$^{-1}$) of the galactic nucleus. The ionized gas has low velocity ($\sim$--50 km s$^{-1}$) and low turbulence ($\sigma_{\rm gas} \lesssim$ 50 km s$^{-1}$), and is kinematically decoupled from the stellar motions, indicating that the gas kinematics are not AGN driven. 
Markarian 950 is the third post-starburst galaxy to host a weak nuclear ionizing source as well as an EELR and a TDE. The overall properties of these three galaxies, including the kinematics and accretion history, are unusual but strikingly similar. We estimate that the incidence of EELRs in PSB-TDE hosts is a factor of $\sim 10 \times$ higher than in other PSB galaxies. This suggests that a gas-rich post-merger environment is a key ingredient in driving elevated TDE rates.
Based on the current observations, we cannot rule out that the EELRs may be powered through an elevated TDE rate in these galaxies. 
If the EELRs are not TDE-powered, the presence of intermittent AGN activity, and in particular the fading of the AGN, may be associated with an increased TDE rate and/or an increased rate of detecting TDEs. 
\end{abstract}
\keywords{Tidal disruption events --- accretion, accretion disks --- black holes --- galaxies: kinematics and dynamics}

\section{Introduction}
\label{sec:introduction}
The tidal disruption of stars by supermassive black holes (SMBHs) is a natural outcome of many-body dynamics in galactic nuclei. Such tidal disruption events (TDEs) occur when a star reaches an orbital pericentre where the stellar self-gravity becomes smaller than the tidal forces exerted by the SMBH. As a result, the star is destroyed and the bound material resulting from the encounter (roughly half of the stellar debris) eventually triggers the process of accretion (e.g. \citealt{Rees1988}). Such events should happen in both quiescent and active galactic nuclei (AGN), although their identification may be much harder in the latter due to the presence of a broad/narrow line region and/or a dusty, obscuring torus. The current sample of known TDEs is biased (to an unknown degree) to TDEs in inactive galaxies by selection, as spectroscopic follow-up surveys tend to avoid AGN host galaxies. 

TDE host galaxies are known to have a strong over-representation among post-starburst (E+A) galaxies \citep{French17, Law-Smith17, Graur18}. These galaxies comprise $<$1 \% of the local galaxy population. They have experienced a recent ($\lesssim$ Gyr ago) starburst, followed by a rapid quenching episode (on timescales of 100--200 Myrs) that leads to their characteristic properties. These include a significant population of young A-type stars as well as low levels of current star formation (although some of it may be obscured, \citealt{Baron23}). TDE hosts are often found to reside in the green valley \citep{Hammerstein21} and are centrally concentrated \citep{Graur18}, suggestive of a short-lived transition phase between star forming and quiescent galaxies also inhabited by PSB galaxies.  
Taken together, this may point towards a connection between the post-starburst state of the galaxy and an elevated rate of star--SMBH encounters. Several mechanisms have been proposed to explain (or at least contribute to) this connection, including disturbed potentials and stellar orbits induced by SMBH binaries (e.g. \citealt{Chen2011}), elevated central density profiles as a result of recent starbursts \citep{Stone2016, Stonevanvelzen2016, Bortolas22}, circum-nuclear gaseous disks \citep{Kennedy16}, and eccentric stellar disks \citep{Madigan2018}. \citet{Stone2018} argued that based on the observed delay time distribution of TDEs (probed by the post-starburst age of the host galaxies), the SMBH binary scenario cannot be the dominant mechanism while stellar overdensities and radial velocity anisotropies can explain the observations for reasonable parameters.

Most of these scenarios are expected to involve an AGN phase at some point in the (post-merger) evolution of the galaxy. Characterizing the population of TDEs along the post-merger sequence (including the starburst and AGN phases) is important to better understand which, if any, of the aforementioned mechanisms is most effective at increasing the rate of TDEs. 
\citet{Prieto16} found that the post-starburst host galaxy of the TDE ASASSN--14li hosted a weak AGN and an extended emission line region (EELR). Such EELRs can be used to trace the accretion history of the galactic nucleus over $\sim$10$^4$ yr timescales. A large sample study of EELRs in 93 PSB galaxies by \citet{French23} yielded six sources with EELRs. Intriguingly, one of these is the host galaxy of the TDE AT2019azh \citep{Hinkle21}, and another is the host galaxy of the quasi-periodic X-ray eruption (QPE) source RXJ1301 \citep{Giustini20}. 

Motivated by these results, we investigated publicly available MUSE (spatially resolved spectroscopy) datasets of 16 TDE host galaxies. In this paper we report on the discovery of a new EELR in the post-starburst galaxy Markarian 950 (Mrk 950 herafter), located at a redshift of $z=0.01628$ which corresponds to a luminosity distance of 66 Mpc, assuming a $\Lambda$-CDM cosmology with H$_0$ = 69.6 km s$^{-1}$ Mpc$^{-1}$, $\Omega_M = 0.29$ and $\Omega_{\Lambda} = 0.71$ \citep{Fixsen96}. This galaxy hosted the TDE iPTF--16fnl, and is the third post-starburst TDE host galaxy to also have an observable EELR. We characterize the properties of Mrk950 in Section \ref{sec:analysis}, and present the results of our analysis in Section \ref{sec:results}, where we discuss them in the context of EELR-hosting PSB+TDE host galaxies and the mechanism responsible for the increased TDE rates.
Our main results are summarized in Section \ref{sec:summary}.

\section{Galaxy properties and new observations}
\label{sec:muse}
An environment search in the NASA/IPAC extragalactic database reveals that Mrk950 has three potential physical companions with heliocentric velocities within 500 km s$^{-1}$ and located at distances of 300--550 kpc. This suggests that it may be part of a small group of galaxies.
In addition to being a post-starburst galaxy, it is also the host galaxy of the transient iPTF--16fnl. 
This transient was identified as a TDE based on the presence of bright UV emission, in combination with broad Balmer, He\,\textsc{ii} and Bowen emission lines that evolved over time \citep{Blagorodnova17, Onori19}. Mrk950 has been modeled and studied previously \citep{Blagorodnova17, Onori19} in the context of the transient TDE emission. \citet{Wevers17} used late-time X-shooter (long-slit) spectroscopy to measure a velocity dispersion of 55$\pm$2 km s$^{-1}$, indicating a low mass SMBH (M$_{\rm BH}$ = 10$^{5.5 \pm 0.42}$ M$_\odot$). \citet{Blagorodnova17} model the host galaxy spectrum and classify it as an E+A galaxy with a single burst of star formation that occurred 650$\pm$300 Myr ago. Based on the line ratios measured in the late-time X-shooter data, \citet{Onori19} conclude that the galaxy may harbor a weak AGN. 

Mrk 950 was observed by the Multi Unit Spectroscopic Explorer \cite[MUSE,][]{Bacon_2010} on 2021 August 16 (MJD 59442), 1813 days after the TDE was discovered. At this epoch the host galaxy emission completely dominates the data (e.g. \citealt{Onori19}). MUSE is an integral field spectrograph providing spatially resolved spectroscopy for each of $\sim$100\,000 spaxels in its 1$\times$1 arcmin field of view. The spectral resolution of MUSE is $\approx$ 2.6 \AA\ (R$\sim$3000), corresponding to a velocity resolution of 160 (80) km s$^{-1}$ at the red and blue ends of the wavelength range, respectively. Two 600 second exposures were obtained in wide-field mode with the adaptive optics facility (AOF), at an average airmass of 1.866. The ambient seeing conditions were $\sim$1.0 arcsec; we estimate the delivered spatial resolution as measured from point sources in the field of view (i.e. including the corrections from the AOF) to be $\sim$0.9 arcsec around 5000\AA\ and $\sim$0.8 arcsec around H$\alpha$. The typical uncertainty in the absolute flux calibration of the pipeline reduced data products is 4--7\% \citep{Weilbacher20}.
\begin{figure*}[!ht]
    \centering
    \includegraphics[width=0.5\textwidth]{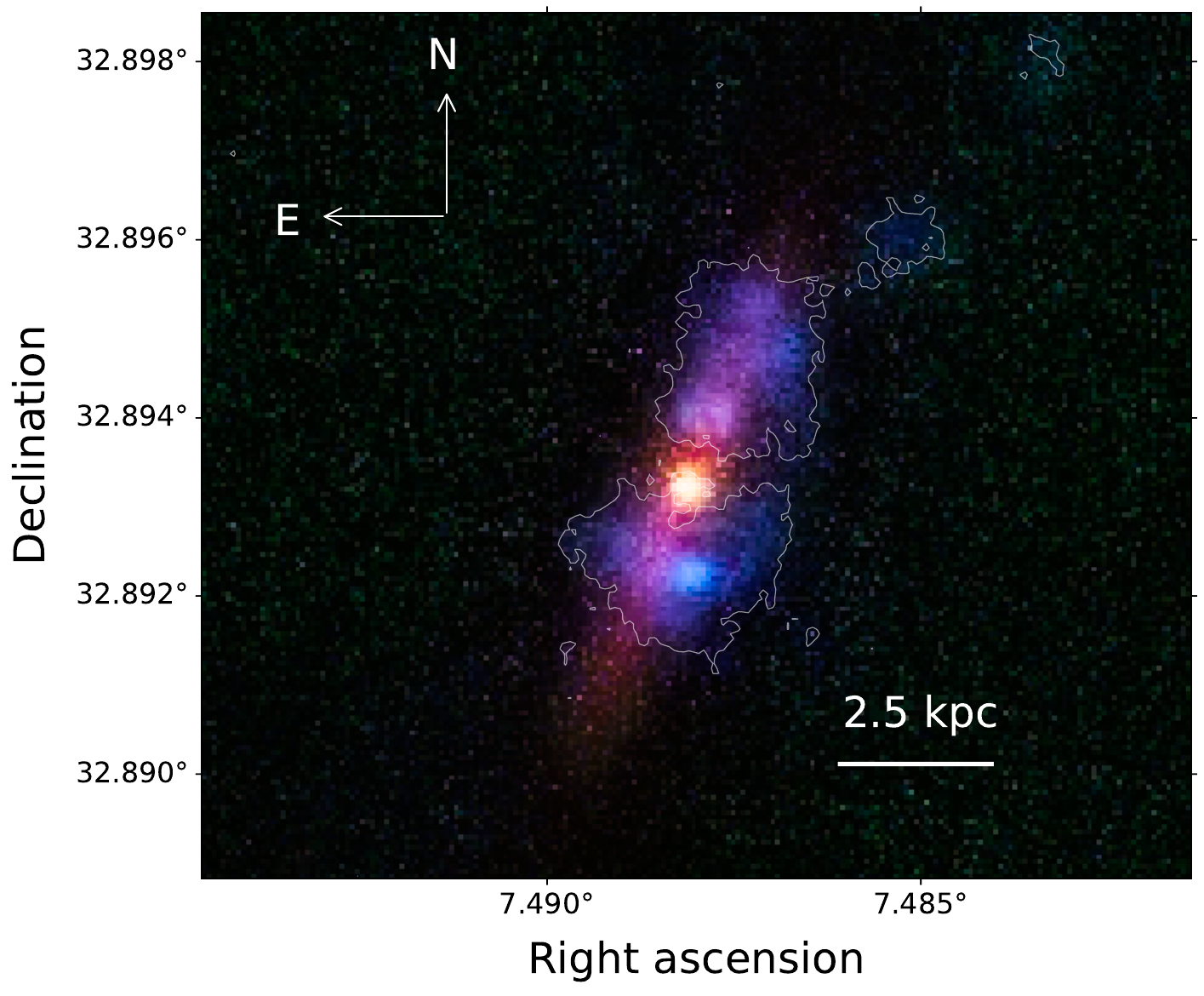}
    \includegraphics[width=0.49\textwidth]{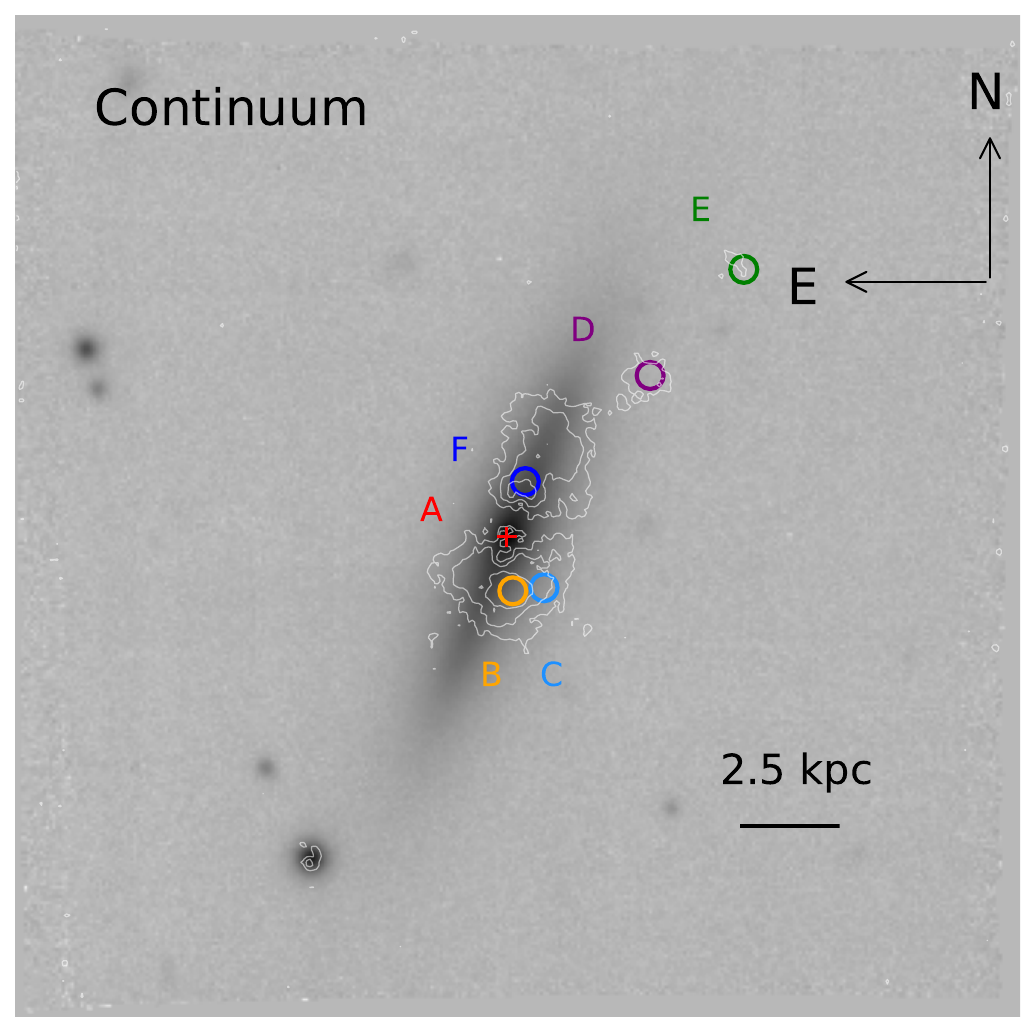}
    \caption{{\bf Composite color image and emission line contour map}. Left: three color composite image combining the continuum emission (7200--7500 \AA, in red), H$\alpha$ (green hues) and [O\,\textsc{iii}] (blue hues) emission. These maps reveal spatially extended emission line regions far beyond the stellar continuum. Note that this panel does not cover the full MUSE field of view, only the regions of interest. Right: emission line contour maps of [O\,\textsc{iii}] overlaid on a continuum image. A red cross marks the center of the continuum emission for reference. Several regions identified by colored circles are used for detailed spectral analysis. The contours in the right panel represent an [O\,\textsc{iii}] flux level of 60 (outer), 150 (middle) and 300 (inner) $\times$ 10$^{-20}$ erg cm$^{-2}$ s$^{-1}$ \AA. The lowest flux contour is also shown in the left panel.}
    \label{fig:contours}
\end{figure*}

\section{Analysis}
\label{sec:analysis}
After downloading the standard pipeline products from the ESO science archive, we correct the datacube for Galactic interstellar extinction, assuming an E(B-V) = 0.07 \citep{Schlafly2011} and the Galactic extinction law of \citet{Fitzpatrick19}. Then we make narrow-band emission line maps and a continuum image to study the gas and stars in the system.

\subsection{Morphology}
We study the gas through the four main emission lines that are required to produce diagnostic diagrams: H$\alpha$, H$\beta$, [O\,\textsc{iii}] and [N\,\textsc{ii}]. [S\,\textsc{ii}] is very weak, hence we do not study it in detail.

\begin{figure*}
    \centering
    \includegraphics[width=\textwidth]{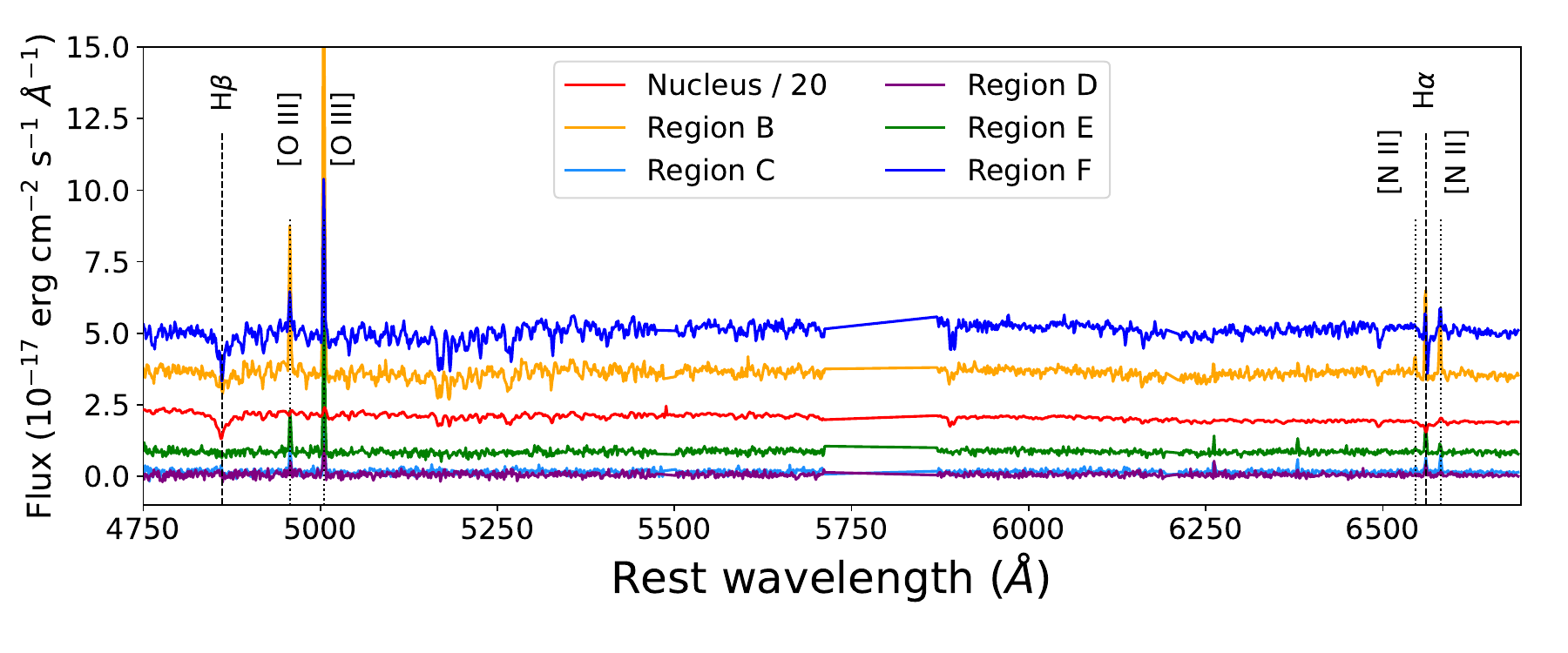}
    \includegraphics[width=0.8\textwidth]{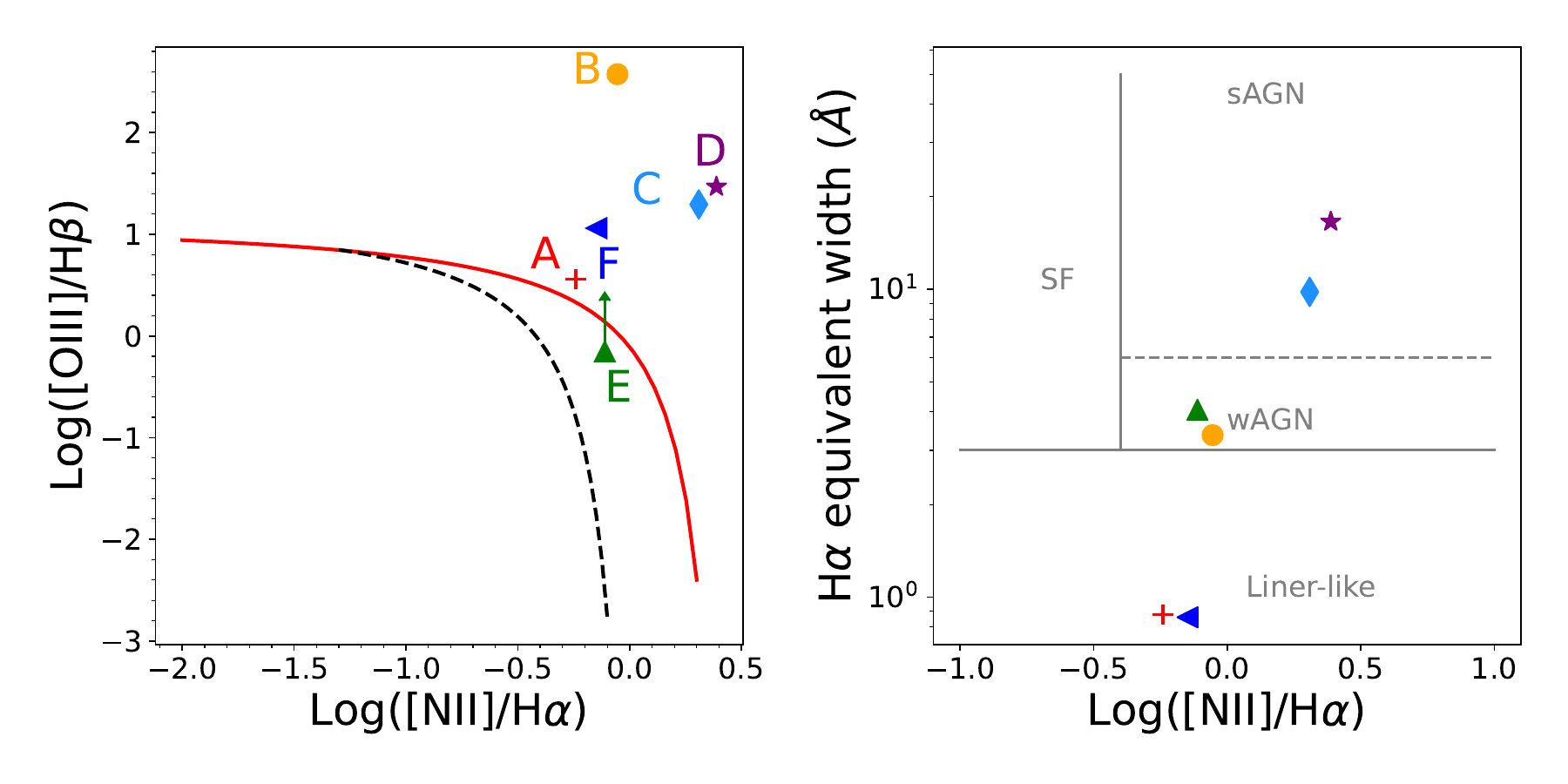}
    \caption{{\bf Spectra and diagnostic diagrams.} Top: Spectra extracted from the regions indicated in Figure \ref{fig:contours}, illustrating the lack of continuum emission and the emission lines in regions C, D and E. The nuclear spectrum (red) has been rescaled by a factor of 20 for clarity. 
    Bottom: BPT (left) and WHAN (right) diagnostic diagrams of these aperture spectra. A hard/non-stellar ionizing continuum is required to power the extended emission, although the nucleus does not currently host a powerful AGN (red cross in the LINER-like region). }
    \label{fig:aperturespectra}
\end{figure*}

A three color composite image including the continuum (7200--7500\AA, red), H$\alpha$ (green hues) and [O\,\textsc{iii}] (blue hues) is shown in Figure \ref{fig:contours} (left panel). The [O\,\textsc{iii}]$\lambda 5007$ is unaffected by stellar absorption, making it a good tracer to study the spatial morphology without relying on a subtraction of the stellar light, which may introduce biases. 
The H$\alpha$ and H$\beta$ lines are observed in absorption where the light is dominated by the stellar component of the galaxy (e.g. Figure \ref{fig:aperturespectra}). Bright extended emission line regions (EELRs) are obvious in the [O\,\textsc{iii}]$\lambda$5007 line, with very faint emission up to $\sim$6.5 kpc to the North-West of the nucleus. The right panel of Figure \ref{fig:contours} shows a continuum grayscale image where the [O\,\textsc{iii}] emission line contours are overlaid for clarity. 
The EELR spans $\sim$2.5 kpc on either side of the nucleus (region A, marked by a red cross). Further out, there are filaments/clumps of ionized gas at larger distances (e.g. regions D and E). These features appear more patchy and disconnected from the nucleus, whereas the aligned component (in projection) is more compact and more closely traces the galaxy disk plane. Example pseudo-aperture spectra for the indicated regions are shown in Figure \ref{fig:aperturespectra}.

\subsection{Stellar and ionized gas kinematics}
We use the the penalized pixel fitting routine PPXF \citep{Cappellari23} to constrain the stellar population properties as well as the kinematics of the stars and the ionized gas. To this end we use the MILES single stellar population library \citep{Vazdekis10, Falcon2011}, which is well matched to the spectral resolution of MUSE, in PPXF. 
We measure the post-starburst age with the X-shooter spectrum from \citet{Onori19} using Bagpipes \citep{Carnall2018, Carnall2019}, assuming a star formation history consisting of an old delayed exponential and a young burst exponential, using a similar method as  \citet{French17, French23} for ASASSN-14li and AT2019azh. The process we use here represents an improvement upon this previous work for ASASSN-14li and AT2019azh, as Bagpipes allows us to fit the entire spectrum, simultaneously modeling the correlated flux uncertainty with three Gaussian process terms \citep{Carnall2019}. The post-starburst age (age since 90$\%$ of the burst stars formed) is $624 \pm 10$ Myr.

The kinematics are shown in Figure \ref{fig:velocitymaps}. 
Measurements on spatially integrated spectra have typical velocity dispersions of $\sim$50--60 km s$^{-1}$ (as measured from the MUSE data in this work as well as X-shooter data, \citealt{Wevers17}), while the spatially resolved stellar velocity dispersion map peaks at values of $\sim$85 km s$^{-1}$. 
The stellar kinematics appear orderly with velocities increasing up to $\sim$100 km s$^{-1}$ along the stellar disk, indicative of the rotational motion of the galaxy. 

The gas kinematics are constrained through a joint fit to the available emission lines (H$\alpha$, H$\beta$, [O\,\textsc{iii}] and [N\,\textsc{ii}]$\lambda$ 6584) in each spaxel\footnote{Decoupling the emission lines yields consistent kinematics.}. In the nucleus the line velocities are low ($<$ 25 km s$^{-1}$), consistent with the motion of the stars. In the EELRs, however, the line velocities are consistently blueshifted by --40 to --70 km s$^{-1}$. This is the case for both the inner part of the EELR (regions B, C, F) as well as the blobs of ionized gas further out (regions D, E). Emission line velocity dispersions range from $\sim$125 km s$^{-1}$ in the nucleus, to 50 km s$^{-1}$ or less in the EELRs, suggesting that there is very little turbulent motion in the extended ionized gas. 

\subsection{Diagnostic diagrams}
We compute line ratios to locate the various selected regions on the diagnostic Baldwin-Philips-Terlevich (BPT) diagram \citep{Baldwin1981}, shown in Figure \ref{fig:aperturespectra}. The dashed and solid red lines can be used to determine the dominant ionization mechanism (e.g. \citealt{Kewley2006}): below the dashed line, galaxies are expected to be star formation dominated \citep{Kauffmann2003}, while the red (extreme starburst) line of \citet{Kewley2001} indicates the maximum line ratios that can be produced by ionization models based on stellar atmospheres; above this line, a harder (non-stellar) continuum is required. In between these two lines is a {\it composite} zone where both SF and AGN can produce the line ratios.

The high SNR aperture spectra are all indicative of a hard ionizing continuum, inconsistent with star formation. This is further corroborated by the WHAN diagram \citep{Cid2011}, where the nucleus is in the LINER-like / retired portion of the diagram, whereas the regions in the EELR indicate a non-stellar ionizing continuum. We include a reddening component in fitting these spectra and find typical values of E(B--V)$\sim$0.5. The Balmer decrements of the emission lines are consistent with theoretical expectations for case B recombination (H$\alpha$/H$\beta$ = 2.86), indicating that additional extinction is unlikely to strongly bias these measurements. 

\subsection{EELR energy budget}
\begin{figure}
    \centering
    \includegraphics[width=\columnwidth]{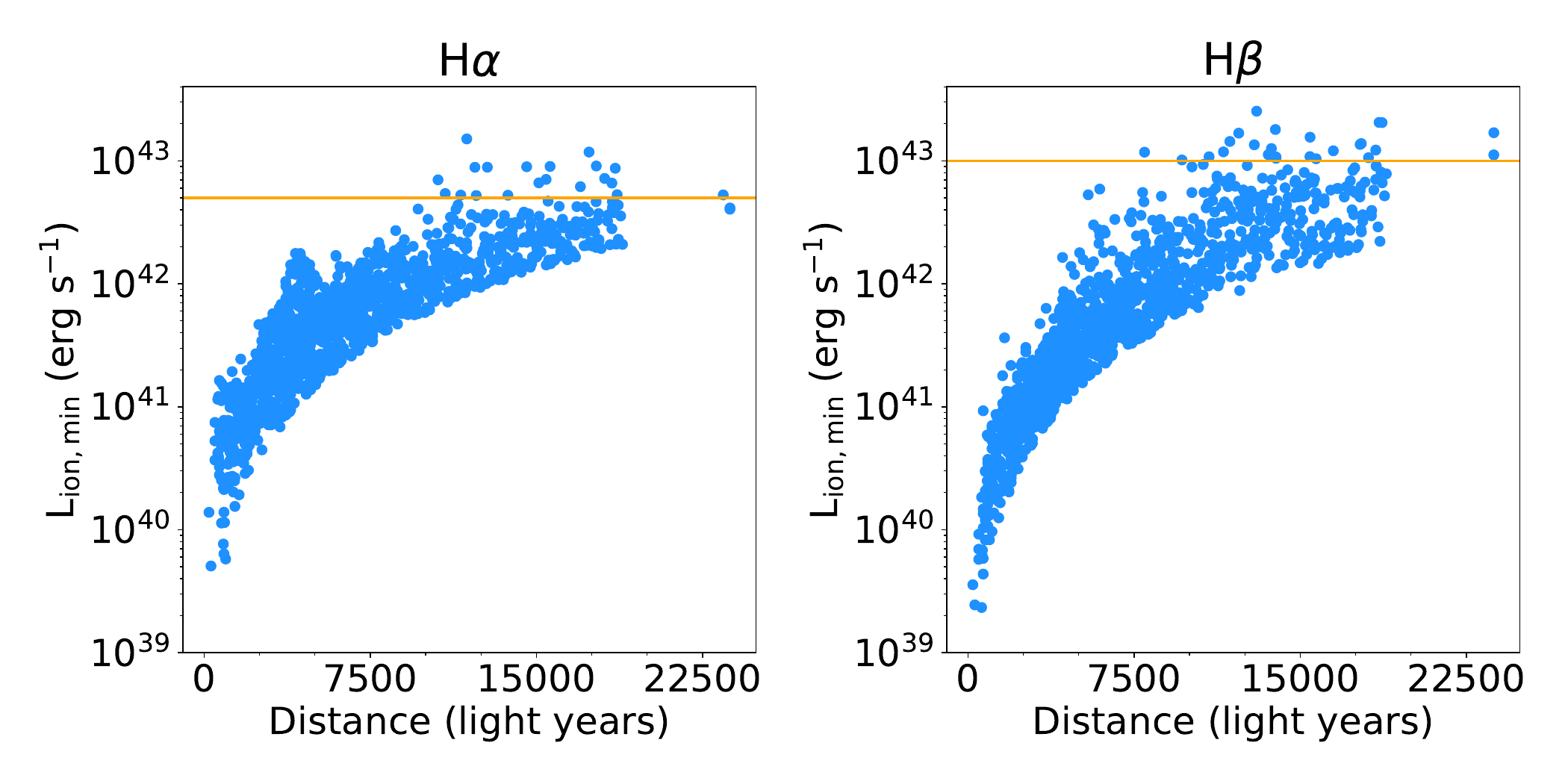}
    \caption{{\bf Constraints on the ionizing luminosity.} Estimate of the nuclear ionizing luminosity required to power the recombination H Balmer lines in the EELR. The upper envelope traces the lower limit of L$_{\rm ion, min}$ (marked by an orange line). We have removed low SNR spaxels with line fluxes $<$25$\times$10$^{-20}$ erg cm$^{-2}$ s$^{-1}$.}
    \label{fig:lion}
\end{figure}

An estimate of the minimum energy required to power the observed narrow line emission (i.e. trace the history of the nuclear engine luminosity) can be obtained by using recombination lines \citep{Keel12, Keel17, French23} because the recombination time ($\sim$100 years) is short compared to the light travel time ($>$10\,000 years). Assuming that the observed H Balmer emission is in recombination balance provides a lower limit on the ionizing luminosity. Parcels of gas at progressively larger distances require larger ionizing luminosities and thereby provide the most stringent constraints. 
We follow \citet{French23} to compute L$_{\rm ion, min}$ from both the H$\alpha$ and H$\beta$ emission line luminosities, $L_H$, in each spaxel, measured after correcting for the stellar component: 
\begin{equation}
    L_{\rm ion, min} = \frac{n_r L_H E_{\rm ion}}{E_H} \frac{1}{f}
\end{equation}
Here, $n_r$ is the number of ionizing photons per recombination, $E_{\rm ion}$ is the ionizing continuum energy between 13.6 and 54.6 eV (i.e. assuming that higher energy photons will be absorbed primarily by Helium), $E_H$ is the energy per H$\alpha$ or H$\beta$ photon, and f is the covering fraction in units of spaxels, 
\begin{equation}
   f = \frac{(2\ \rm arctan(0.5/r))^2}{4 \pi}
\end{equation}
The distribution of values with galactocentric radius is shown in Figure \ref{fig:lion}. From the H$\beta$ line we find that L$_{\rm ion, min} \gtrsim $ 10$^{43}$ erg s$^{-1}$, while the H$\alpha$ estimate yields L$_{\rm ion, min} \gtrsim $ 5 $\times$10$^{42}$ erg s$^{-1}$.
Note that because of the relatively low SNR of individual spaxel spectra we have not included an extinction component to estimate these H Balmer line fluxes (while a correction was included for the aperture spectra), and hence these are conservative estimates as any additional extinction correction would increase the line fluxes and the inferred L$_{\rm ion}$.
For the typical E(B--V) values inferred from the aperture spectra, this correction may be a factor of $\sim$2--4.

\subsection{Nuclear luminosity estimates}
We measure (SSP- and extinction-corrected) nuclear integrated luminosities (using an aperture with radius = 0.9 arcsec) of L$_{\rm H\alpha, nuc} =$ 1.8 $\times$ 10$^{38}$ erg s$^{-1}$, L$_{\rm H\beta, nuc} =$ 8 $\times$ 10$^{37}$ erg s$^{-1}$ and L$_{\rm [O\,\textsc{iii}], nuc} =$ 2.3 $\times$ 10$^{38}$ erg s$^{-1}$. 
Using bolometric corrections from \citet{Kauffmann09} and \citet{Netzer09} for the [O\,\textsc{iii}] and H$\beta$ lines respectively yields estimates of L$_{\rm AGN, bol}$ = 1--2.5$\times$10$^{41}$ erg s$^{-1}$ for the present-day AGN luminosity. As a more conservative probe, from the nuclear IR luminosity we estimate that L$_{\rm IR}$ $<$ 4$\times$10$^{42}$ erg s$^{-1}$ based on IRAS all-sky survey upper limits (see e.g. \citealt{Baron22}). We correct this estimate for low level star formation by using the calibration of \citet{Wild2010}, based on the distance between the source and the locus of the SF main sequence in the BPT diagram, to estimate the contribution of star formation. This correction amounts to $\sim$30 per cent, leading to L$_{\rm IR,AGN}$ $<$ 2.5$\times$10$^{42}$ erg s$^{-1}$.

\section{Results and discussion}
\label{sec:results}
We have discovered extended emission line regions in [O\,\textsc{iii}], H$\alpha$ and [N\,\textsc{ii}] in the post-starburst, TDE host galaxy Mrk950. The availability of MUSE IFU data facilitates a detailed study of properties including morphology, kinematics and ionization mechanism. 

While absolute stellar ages are subject to significant uncertainties, their spatial distribution can be inferred to establish a (relative) trend. The SSP age analysis indicates that a young ($<$1 Gyr) stellar population is preferred at the center of the galaxy, and the SSP age increases with galactocentric radius. This is consistent with the post-starburst nature, where centrally concentrated starbursts are often seen \citep{Yang2008}; alternatively this could also be due to outside-in quenching \citep{Chen2019}. We find a light-weighted age in the nucleus of $\sim$10$^{8.75} \approx 5\times10^{8}$ yr, which is subject to significant systematic uncertainties due to the lack of wavelength coverage blueward of H$\beta$ but consistent with an independent estimate of the time since the starburst (650$\pm$300 Myr, \citealt{Blagorodnova17}).
\begin{figure*}[!t]
    \centering
    \includegraphics[width=\textwidth]{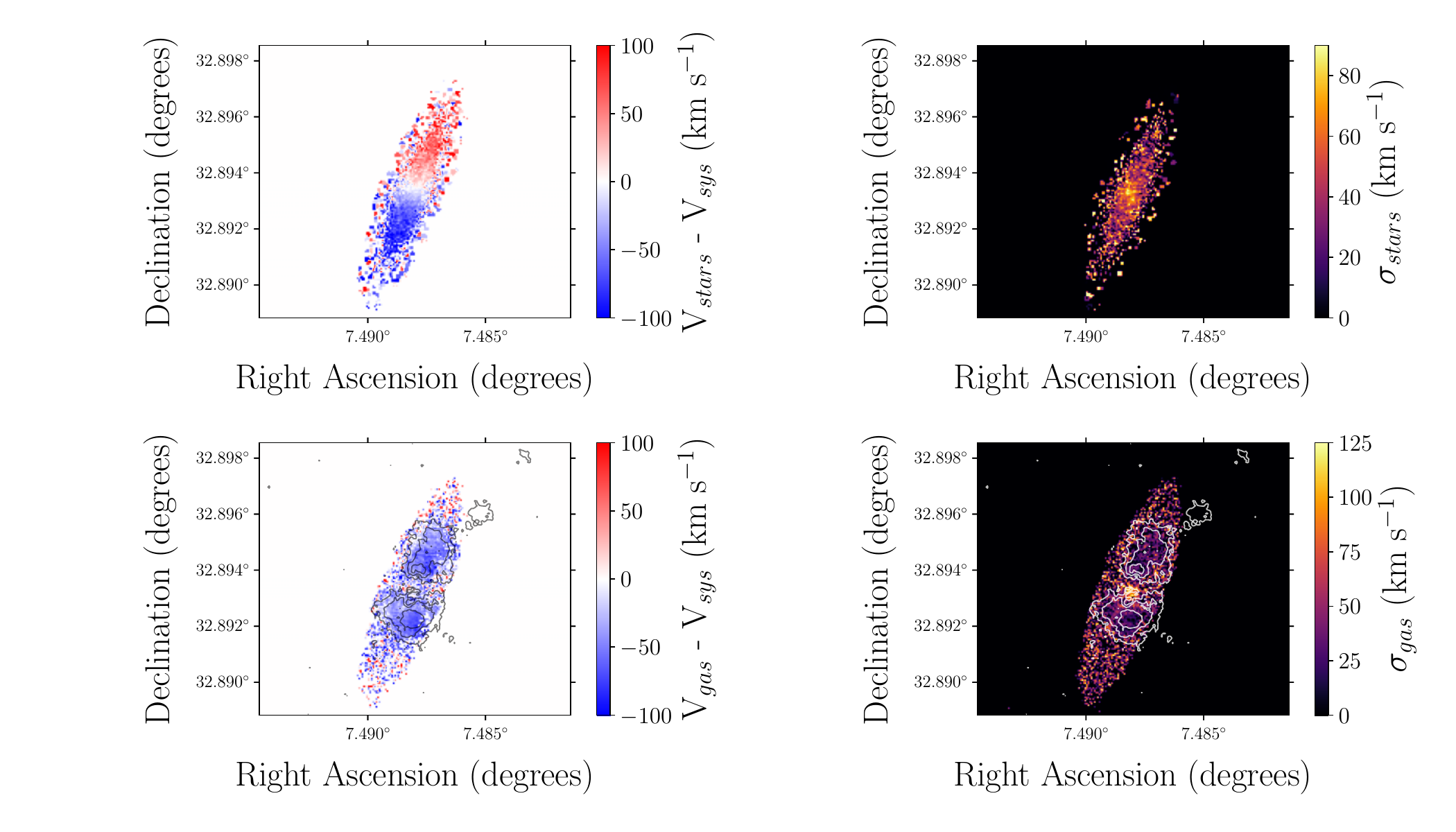}
    \caption{{\bf Stellar and gas kinematics.} Top: stellar velocity (left) and velocity dispersion (right). Stellar rotation is evident, and the stellar velocity dispersion peaks in the nucleus. Bottom: gas velocity (left) and velocity dispersion. The velocity of the gas is consistently blueshifted irrespective of the stellar motions, indicating that the gas is decoupled from the stars. The gas velocity dispersion peaks at a higher value than the stellar dispersion in the nucleus, while it is very low across the EELR. }
    \label{fig:velocitymaps}
\end{figure*}

\subsection{Accretion/merger history}
The stellar component of the galaxy appears rotation-dominated (Figure \ref{fig:velocitymaps} top left), with a stellar velocity dispersion that increases towards the center (Figure \ref{fig:velocitymaps} top right). This is in stark contrast with the ionized gas kinematics (bottom panels of the same figure), which show a consistent blueshift across the disk of the galaxy and beyond. In other words, the ionized gas kinematics are completely decoupled from the stellar motions. The consistency between the inferred velocities in all parts of the EELR (blueshifts of order --50 km s$^{-1}$) suggests that while the gas is decoupled from the stars, the various ionized gas components are kinematically connected. The very narrow velocity dispersion of the emission lines indicate that none of the gas, save for that located in the nucleus, is turbulent. This implies that the AGN is currently too weak to drive the uniform gas motion through radiative pressure, which is consistent with the lack of broad emission lines and a flat IR color. The low intrinsic line broadening, comparable to the sound speed at $\sim$10\,000 K, also disfavors shocks as a feasible excitation mechanism. 

We propose that the ionized gas is located in the foreground with respect to the nucleus, and we are seeing a relic ionization cone powered by a (currently weak) nuclear source that is pointed along our line of sight. The EELR gas was likely deposited at large distances from the galaxy as the result of a gas-rich merger. Tidal tails/merger features can last for hundreds of Myrs \citep[e.g.,][]{Pawlik2016}, consistent with the inferred time since starburst. This can also help to explain the more isolated (in projection) patchy gas with similar kinematics at larger distances (4--6.5 kpc), as a result of splash-back following a gas-rich merger. 
Nevertheless, diagnostic diagrams indicate that the ionization state, save for the nucleus, is powered by a non-stellar continuum as the majority of the EELR is located beyond the extreme starburst line. 

\subsection{Similarities among TDE host galaxies hosting EELRs}
A detailed comparative analysis of the stellar and ionized gas properties, including morphology, stellar ages, and kinematics, reveal a striking number of similarities shared between the three EELR-hosting TDE host galaxies. These properties are summarized quantitatively in Table \ref{tab:tdehosts}.
\begin{enumerate}
    \item All three galaxies have a low mass ($\sim 10^{6}$ M$_{\odot}$) black hole \citep{Wevers17, Wevers20} and are post-starburst (E+A) \citep{Prieto16, Blagorodnova17, Hinkle21}.
    \item All three galaxies host very extended ($\gtrsim$ 15\,000 light years) emission line regions. The MUSE data for ASASSN--14li and iPTF--16fnl reveal a patchy morphology with clumps of gas at large projected separations. 
    \item The EELRs have very narrow line velocity dispersion (20--65 km s$^{-1}$) and low velocities, inconsistent with their kinematics being AGN-driven (outflows) or ionization by fast shocks. The most likely origin for the copious amounts of gas in these galaxies is therefore through gas-rich mergers. The lack of visual disturbances in the continuum emission and prominent dust lanes suggests that these were minor mergers.
    \item The line fluxes and ratios indicate that the EELRs are photo-ionized by a continuum source harder than any stellar continuum can provide. Moreover, the current luminosity of the nuclear engine is significantly fainter than required to explain the emission line fluxes for at least 2/3 sources (for ASASSN--14li the IRAS upper limits are not constraining). This discrepancy cannot be explained through excess extinction and requires a physical origin.
\end{enumerate}
The similarity in properties suggests that the underlying mechanism driving the observed preference is the same for all sources.
\begin{table*}
\caption{Summary of the TDE host galaxy properties with an observed EELR. All three hosts are post-starburst galaxies. Black hole masses are based on velocity dispersion measurements. $\sigma_{\rm gas}$ denotes the typical velocity dispersion in the EELR.} 
    \centering
    \begin{tabular}{ccccccccc}
    TDE & Resolution & log$_{10}$(M$_{\rm BH}$) & log$_{10}$(L$_{\rm IR,AGN}$) & log$_{10}$(L$_{\rm ion, min}$) &  PSB age & $\sigma_{\rm gas}$ & BPT+WHAN & BPT+WHAN  \\
         &  (pc / pix) & (M$_{\odot}$) & (erg s$^{-1}$) & (erg s$^{-1}$) &  (Myr) & (km s$^{-1}$) & (EELR) & (nucleus) \\\hline
         iPTF--16fnl &  75 & 5.5$\pm$0.4 & $<$42.4 & $>$43.0 & 624$\pm$10 & 50 & AGN & LINER \\
         ASASSN--14li & 84 & 6.2$\pm$0.4 & $<$42.3 & $>$41.3 & 415$^{+370}_{-70}$ & 40 & AGN & LINER   \\
         AT2019azh &  224 & 6.4$\pm$0.4 & $<$42.1 & $>$42.9 & 200$\pm$30 & 60 & AGN & LINER 
    \end{tabular}
    \label{tab:tdehosts}
\end{table*}

\subsection{Quantifying the over-representation of TDEs among EELR-hosting galaxies}
iPTF--16fnl / Mrk 950 is the third post-starburst TDE host galaxy hosting an EELR observed with an IFU, following ASASSN--14li \citep{Prieto16} and AT2019azh \citep{French23}. We inspected a total of 16 additional MUSE cubes of TDE host galaxies (including the three aforementioned sources), and found no clear evidence for EELRs in other hosts. 
The implication is that the EELR incidence in all TDE hosts (0.19$^{+0.14}_{-0.11}$, where the uncertainties denote the two-sided 68\% confidence interval for Poisson statistics) is elevated by a factor of $\sim$2--5 compared to results from larger galaxy studies \citep{Sanchez18, Lopez-Coba_2020, Keel24}. The true over-representation may be higher, as the majority of sources in sample studies show evidence for AGN-driven kinematics (velocity dispersions $>$100 km s$^{-1}$ and outflows) rather than the low velocity, non-turbulent gas seen in these TDE hosts.

Considering only the 5 sources in the TDE host MUSE sample that are classified as post-starburst galaxies, the fractional incidence of EELRs is 0.6$^{+0.4}_{-0.33}$. This is an order of magnitude higher than the fraction of EELRs found in a sample of 93 post-starburst galaxies (0.065$^{+0.035}_{-0.025}$) reported by \citet{French23}. 

The over-representation of TDE hosts among PSB galaxies has been studied in detail \citep{Arcavi2014, French17, Law-Smith17, Graur18, French20}. The overabundance of EELR-hosting PSB galaxies by an order of magnitude among TDE hosts in combination with their strikingly similar properties strongly suggests a connection between the TDE rate enhancement and this (very short-lived) gas-rich phase in post-merger evolution. 

\subsection{What powers the EELRs in PSB-TDE host galaxies?}
The implications of this over-representation depend on the assumed powering mechanism of the EELRs. Implicitly this is assumed to be an AGN, but we will show below that the current constraints do not rule out TDEs as the source of the energy required to power the EELRs. 

\subsubsection{AGN-powered EELRs}
In the scenario that the EELRs are AGN powered, we can estimate the present-day AGN luminosity and compare it to the ionizing luminosity required to power the observed EELR. 
The present-day AGN luminosity estimates based on the emission lines are lower than the minimum luminosity required by both the H$\alpha$ and H$\beta$ lines by a factor of $\sim$50--100. The more conservative limit obtained from the IR luminosity is also lower than required by a factor of $\gtrsim 4$. The implication is that if the EELR is AGN powered, it must have faded (by a factor of at least a few, and potentially up to 2 orders of magnitude) in the recent past. A similar situation occurs for AT2019azh \citep{French23}, while for ASASSSN--14li the IR upper limits are not constraining enough (Table \ref{tab:tdehosts}).

In this scenario, TDEs appear to select a very specific and short-lived window along the post-starburst/post-merger phase, where a putative AGN recently had a significant decrease in its luminosity output. The EELRs do not appear to have a clear inner boundary, implying that the fading of the putative AGN happened within the last $\sim$1000 years (which is the typical spatial/PSF FWHM image quality at the host redshifts). We speculate that a sufficiently elevated TDE rate may affect the AGN accretion disk / nuclear gas inflows through feedback processes (e.g. \citealt{Wevers24}), which may prevent the circumnuclear gas from feeding the accretion disk and help to explain the sudden decrease in AGN luminosity, the lack of broad emission lines, and the overrepresentation of EELRs in fading AGNs among TDE hosts. Another possibility is that one or more TDEs trigger significant changes in the AGN accretion flow/structure and (temporarily) decreasing the accretion rate. Such disruptive events may be related to changing-look AGNs, Bowen fluorescence flares or other ambiguous nuclear transients (e.g. \citealt{Trakhtenbrot19, Neustadt20}). Testing whether any of these scenarios are intrinsically related to TDEs may be possible with IFU studies of their host galaxies.

Alternatively, AGN can flicker on the timescales observable with EELRs $\sim10^4-10^5$ years \citep{Schawinski15, Keel2017}, and these timescales are much shorter than the PSB phase. It is therefore not clear whether the phase where we observe PSBs to have EELRs is a single short-lived phase, or a phase that will occur multiple times over the $\sim 10^8-10^9$ years of the PSB phase. With a larger sample of TDE host galaxies exhibiting EELRs and their post-starburst ages, it may be possible to determine whether there is a preferential delay time with respect to the starburst where the TDE rate is elevated. Note that in this scenario, all three sources would have experienced the same sudden decrease in luminosity prior to the TDE detection, which may be a selection effects related to the stochastic nature of typical AGN flickering variability.

\subsubsection{TDE-powered EELRs}
Thus far we have made the implicit assumption that the presence of an EELR is the result of a recent change in the luminosity history of an AGN. 
However, if the TDE rate is high enough they can also inject significant amounts of energy into the interstellar medium.

We can use our IFU observations to try and constrain the local TDE rate required to be consistent with the observed EELRs in each individual galaxy by using the typical spatial resolution of our observations. A lower limit estimate can be obtained from the native pixel size, which is of the order of 100--250 lightyears at the host redshifts (Table \ref{tab:tdehosts}). 
With this spatial resolution, we can only distinguish between TDE {\it light echoes} if they occur more than $\sim$100s of years apart. 
The inferred local rates (few $\times$ 10$^{-3}$ -- 10$^{-2}$ yr$^{-1}$), while elevated compared to the global TDE rate, appear plausible for most of the mechanisms that may be responsible for increasing the TDE rate. 

Higher spatial resolution imaging/IFU observations can be used to improve these constraints by decreasing the spatial scales on which we do not detect shell-like structures separated by the light travel time between subsequent TDEs. Smaller spatial scales will lead to higher inferred rates. Observations with a factor of $\sim$10 better spatial resolution will be able to rule out TDEs as a powering mechanism for the EELRs if no shell-like structures are detected, as they would require local TDE rates incompatible with the sample-averaged TDE rates even for post-starburst galaxies. 

\subsection{Implications for the mechanism driving the TDE rate increase}
We highlight that the rate of EELRs in PSB TDE hosts (3/5) is much higher than the rate of EELRs in general PSB hosts (6/93, \citealt{French23}). This either means that some additional mechanisms are increasing the rate of both TDEs and EELRs in some of the post-starburst galaxies, or that the duty cycles of the AGN and TDE rate enhancement are coupled in some way, as to have the TDE rate peak in the EELR phase moreso than in the long-quiescent-nucleus portions of the PSB phase. 
The short visibility of EELRs can provide stringent constraints on the relative timing, although it is not understood whether galaxies can have multiple short-lived EELR phases during their post-starburst evolution.

It has been proposed that the presence of an AGN accretion disk may increase the TDE rate in a galaxy. \citet{Kennedy16} find that star-disk interactions increase the TDE rate by a factor of $\sim10\times$. \citet{Wang2024} find that the disappearance of an AGN thin disk (i.e. when an AGN becomes inactive) can increase the TDE rate by $\gtrsim100\times$ due to star-disk interactions as well as the presence of stars formed in the outer regions of the accretion disk. While these mechanisms can increase the rate of TDEs, detecting the flares may be more challenging for AGN with on-going accretion, due to selection against AGN variability in transient searches or obscuration from the dusty torus. Predictions for TDEs in AGN disks demonstrate a broad range in observable signatures \citep{Chan2019}, complicating their identification. Although some PSB galaxies may have nuclear dust obscuration \citep{Baron23}, they typically lack the silicate features in NIR spectroscopy often seen in obscured starburst or AGN galaxies \citep{Smercina2018}. For the PSB TDE hosts, central obscuration is unlikely by selection, because we have observed the optical TDE light. PSB galaxies with EELRs may have a high TDE rate because of a combination of physical and observational effects: the flickering AGN disk may enhance the TDE rate while providing enough ``off" time in which the TDE can be correctly characterized, in addition to any other mechanisms increasing the TDE rate in PSB galaxies.

We do note that an increased TDE rate is a natural consequence of large amounts of gas being funnelled to the center of post-merger galaxies. Several scenarios exist which trigger a starburst and subsequently an elevated TDE rate, including SMBH binaries (e.g. \citealt{Stone2011, Chen2011}), high stellar densities \citep{Stone2018}, AGN disks \citep{Kennedy16, Wang2024} and eccentric nuclear stellar disks \citep{Madigan2018}. 
These would all lead to a sharp initial rise in the TDE rate that gradually decreases over time (where this timescale is typically shorter for SMBH binaries than for an eccentric stellar disk). The rate of disruptions may initially be so high (up to $\sim$10$^{-1}$ galaxy$^{-1}$ yr$^{-1}$, consistent with the extreme TDE rates inferred from (ultra-)luminous infrared galaxies, e.g. \citealt{Mattila18}) that the nucleus is indistinguishable from a classic AGN, leading to an observational bias against detecting TDEs in such systems. When the TDE rate decreases sufficiently to allow for the detection of individual events, this would lead to an overabundance of TDE detections compared to the general galaxy population. If the timescale for the TDE rate to decrease is $>$15\,000 years (the typical EELR lifetime), these scenarios can explain most of the peculiar host galaxy preferences of TDEs, including the gas-rich, post-merger and/or PSB preference, the centrally concentrated light profiles, and the overabundance of EELRs. 

Intriguingly, several of these mechanisms would also lead to an increased rate of partial TDEs and extreme mass ratio inspirals in similar host galaxies (e.g. \citealt{Madigan2018, Metzger22}). \citet{French23} found that one of the EELR-hosting PSB galaxies in their sample was a QPE source, which have been suggested to be related to TDEs (e.g. \citealt{Arcodia21, Miniutti23}), perhaps due to repeated partial TDEs (e.g. \citealt{King22}) or interactions between a TDE disk and a pre-existing EMRI \citep{Linial23}. In a companion paper (Wevers et al. submitted) we show that EELRs are also over-represented in QPE host galaxies, and that they have properties that are very similar to the TDE host galaxies with EELRs. 

\section{Summary}
\label{sec:summary}
We have discovered a new extended emission line region following the visual inspection of 16 publicly available MUSE IFU datasets of TDE host galaxies. This brings the total number of EELR-hosting TDE host galaxies to 3 (ASASSN--14li, iPTF--16fnl, and AT2019azh). 
The main new results and implications can be summarized as follows: 
\begin{enumerate}
    \item There is an over-representation of EELRs in TDE-hosting PSB galaxies compared to other PSB galaxies by a factor of $\sim$10. EELR-hosting PSB galaxies make up $<$ 0.1 per cent of the local galaxy population, implying that TDEs are selecting an extremely rare subset of galaxies. 
    \item The TDE host galaxies with an EELR require a non-stellar photo-ionizing source whose current luminosity is incompatible with the EELR energy budget in at least 2/3 sources. These galaxies also have very similar (and peculiar) gas kinematics that are decoupled from the stellar motions. The EELRs are likely the result of gas-rich minor merger events.
    \item We find that in terms of the EELRs being powered entirely by TDEs, the constraints from our IFU observations remain consistent with the PSB TDE rate, i.e., we are not able to rule out this scenario. Future observations with higher spatial resolution can provide better constraints and will be able to decisively test this idea.
    \item If the EELRs are AGN powered, this suggests that intermittent AGN activity, and in particular the recent fading of the AGN, may be coupled to the intrinsic and/or detection rate of TDEs.
    \item Regardless of the exact origin of the ionizing photons, TDEs appear to select a very specific and short-lived window in the post-starburst evolution of (post-merger) galaxies where an EELR is visible. 
\end{enumerate}

In future, narrowing down the mechanism(s) that are responsible for the peculiar host galaxy preference of TDEs could be facilitated by further constraining the nuclear stellar and gas kinematics of their hosts. This can be achieved with IFU observations that have a higher spatial resolution than presented here, to further resolve the cores of these galaxies and enable Jeans modeling to characterize disturbances to their potentials. High spatial resolution imaging of nearby TDEs (e.g. with HST or JWST) can be used to constrain the nuclear stellar density profiles on scales approaching the sphere of influence as well as the presence of asymmetric nuclei, which can help to constrain predictions of the eccentric nuclear stellar disk hypothesis. 

\begin{acknowledgments}
We are grateful to T. Fischer, M. Guolo, K. Rowlands, D. Kakkad, W. Lu, S. van Velzen, and A. Zabludoff for discussions and suggestions, and to J. Depasquale for creating the composite color image in Figure \ref{fig:contours}. We thank the referee for a constructive report that helped to improve the clarity of the paper.
KDF acknowledges support from NSF grant AAG 22-06164.
Based on observations collected at the European Southern Observatory under ESO programme 105.20GS.002 
This research was supported in part by grant NSF PHY-2309135 to the Kavli Institute for Theoretical Physics (KITP). We thank the organizers of the KITP program: Towards a Physical Understanding of Tidal Disruption Events, where this work was completed. 
\end{acknowledgments}
\bibliographystyle{aasjournal}
\bibliography{bib}
\end{document}